\documentclass[twocolumn,aps,superscriptaddress]{revtex4}
\usepackage[latin9]{inputenc}
\usepackage{amsmath}
\usepackage{amssymb}
\usepackage{graphicx}

\makeatletter

\@ifundefined{textcolor}{}
{%
 \definecolor{BLACK}{gray}{0}
 \definecolor{WHITE}{gray}{1}
 \definecolor{RED}{rgb}{1,0,0}
 \definecolor{GREEN}{rgb}{0,1,0}
 \definecolor{BLUE}{rgb}{0,0,1}
 \definecolor{CYAN}{cmyk}{1,0,0,0}
 \definecolor{MAGENTA}{cmyk}{0,1,0,0}
 \definecolor{YELLOW}{cmyk}{0,0,1,0}
}

\usepackage{epsfig}

\begin{document}

\title{A Versatile Source of Single Photons for Quantum Information Processing}

\author{Michael Förtsch}
\affiliation{Max Planck Institute for the Science of Light, Günther-Scharowsky-Str.
1, Bau 24, 91058, Erlangen, Germany}
\affiliation{Institut für Optik, Information und Photonik, University of Erlangen-Nuremberg,
Staudtstraße 7/B2, 91058, Erlangen, Germany}
\affiliation{SAOT, School in Advanced Optical Technologies, Paul-Gordan-Straße
6, 91052 Erlangen}

\author{Josef Fürst}
\affiliation{Max Planck Institute for the Science of Light, Günther-Scharowsky-Str.
1, Bau 24, 91058, Erlangen, Germany}
\affiliation{Institut für Optik, Information und Photonik, University of Erlangen-Nuremberg,
Staudtstraße 7/B2, 91058, Erlangen, Germany}

\author{Christoffer Wittmann}
\affiliation{Max Planck Institute for the Science of Light, Günther-Scharowsky-Str.
1, Bau 24, 91058, Erlangen, Germany}
\affiliation{Institut für Optik, Information und Photonik, University of Erlangen-Nuremberg,
Staudtstraße 7/B2, 91058, Erlangen, Germany}

\author{Dmitry Strekalov}
\affiliation{Max Planck Institute for the Science of Light, Günther-Scharowsky-Str.
1, Bau 24, 91058, Erlangen, Germany}

\author{Andrea Aiello}
\affiliation{Max Planck Institute for the Science of Light, Günther-Scharowsky-Str.
1, Bau 24, 91058, Erlangen, Germany}
\affiliation{Institut für Optik, Information und Photonik, University of Erlangen-Nuremberg,
Staudtstraße 7/B2, 91058, Erlangen, Germany}

\author{Maria V. Chekhova}
\affiliation{Max Planck Institute for the Science of Light, Günther-Scharowsky-Str.
1, Bau 24, 91058, Erlangen, Germany}

\author{Christine Silberhorn}
\affiliation{Max Planck Institute for the Science of Light, Günther-Scharowsky-Str.
1, Bau 24, 91058, Erlangen, Germany}

\author{Gerd Leuchs}
\affiliation{Max Planck Institute for the Science of Light, Günther-Scharowsky-Str.
1, Bau 24, 91058, Erlangen, Germany}
\affiliation{Institut für Optik, Information und Photonik, University of Erlangen-Nuremberg,
Staudtstraße 7/B2, 91058, Erlangen, Germany}

\author{Christoph Marquardt}
\affiliation{Max Planck Institute for the Science of Light, Günther-Scharowsky-Str.
1, Bau 24, 91058, Erlangen, Germany}
\affiliation{Institut für Optik, Information und Photonik, University of Erlangen-Nuremberg,
Staudtstraße 7/B2, 91058, Erlangen, Germany}

\date{\today}

\maketitle

\textbf{The quantum state of a single photon stands amongst the most fundamental and intriguing manifestations of quantum physics \cite{Einstein:1905p7646}. At the same time single photons and pairs of single photons are important building blocks in the fields of linear optical based quantum computation \cite{Knill:2001p5757}  and quantum repeater infrastructure \cite{Sangouard:2011p7243}
. These fields possess enormous potential \cite{Kimble:2008p7640} and much scientific and technological progress has been made in developing individual components, like quantum memories and photon sources using various physical implementations \cite{Lvovsky:2009p7641, Michler:2000p5617,Kurtsiefer:2000p7293,Thompson:2006p6472,Wilk:2007p7297,Lee:2011p7406,Hong:1986p7475}. However, further progress suffers from the lack of compatibility between these different components. 
Ultimately, one aims for a versatile source of single photons and photon pairs in order to overcome this hurdle of incompatibility. 
Such a photon source should allow for tuning of the spectral properties (wide wavelength range and narrow bandwidth) to address different implementations while retaining high efficiency. In addition, it should be able to bridge different wavelength regimes to make implementations  compatible. Here we introduce and experimentally demonstrate such a versatile single photon and photon pair source based on the physics of whispering gallery resonators. A disk-shaped, monolithic and intrinsically stable resonator is made of lithium niobate and supports a cavity-assisted triply-resonant spontaneous parametric down-conversion process. Measurements show that photon pairs are efficiently generated in two highly tunable resonator modes. We verify wavelength tuning over 100~nm between both modes with a controllable bandwidth between 7.2 and 13~MHz. 
Heralding of single photons yields anti-bunching with $g^{(2)}(0) < 0.2$. This compact source provides unprecedented possibilities to couple to different physical quantum systems and makes it ideal for the implementation of quantum repeaters and optical quantum information processing.}

It is known that in a nonlinear medium a photon can spontaneously decay into a pair of photons, usually called signal and idler. This process, referred to as spontaneous parametric down-conversion (SPDC), preserves the energy and momentum of the parent photon. 
The resulting pair of photons posses the ability to bridge different wavelength ranges. At the same time detecting one photon of this pair unambiguously heralds the presence of a single photon. In principle, the process of SPDC has a very high bandwidth. By assisting it with a high quality factor (high-Q) resonator, the desired narrow bandwidth of a few MHz for the individual photons can be ensured \cite{Ou:1999p3986}. 
A thorough description of this resonator-assisted SPDC leads to a two-mode EPR entangled state \cite{Drummond:1990p7084} and has successfully been used to generate heralded single photons \cite{Scholz:2009p4886}. 
Recently, resonator-assisted SPDC has led to a substantial progress towards an efficient narrow-band source \cite{Hockel:2011p4528}. However, the wavelength and bandwidth tunability remained a major challenge.

We overcome this problem by using an optical whispering gallery mode resonator (WGMR). These resonators are based on total internal reflection of light and can be tuned to all wavelengths within the whole transparency range of the material they are made of \cite{Vahala:2003p6874}. WGMRs can have high-Q leading to narrow bandwidths \cite{Ilchenko:2001p7395} that match those of atomic transitions. Therefore, further spatial and spectral filtering \cite{NeergaardNielsen:2007p7081,Halder:2008p6937,Wolfgramm:2011p7029}, which can drastically reduce the systems efficiency, becomes unnecessary. Recently it was shown that the strong second order nonlinearity in lithium niobate can be used in monolithic WGMRs for highly efficient nonlinear processes \cite{Ilchenko:2004p7213,Furst:2010p1048,Furst:2010p6510,Beckmann:2011p6321,Furst:2011p5271}.

\begin{figure}
\begin{tabular}{l}
\centerline{\includegraphics[width=8.6cm]{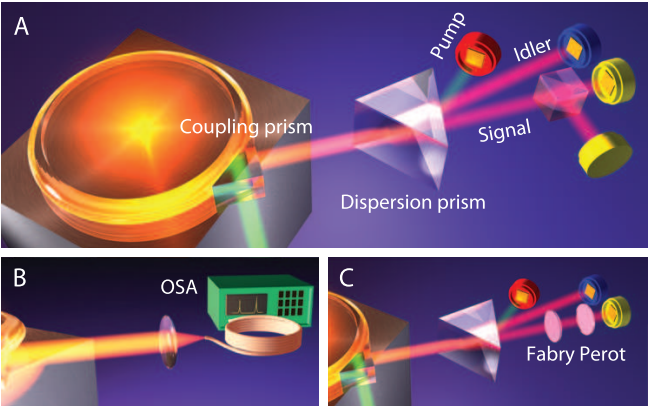}}\\[-0.1cm]
\end{tabular}
\caption{\label{fig.setup} Sketch of the experiment. (A) The whispering gallery mode resonator (manufactured from lithium niobate, radius of 1.9~mm) is pumped by a frequency doubled Nd:YAG laser emitting continuous wave light at a wavelength of $532\,\mbox{nm}$. A diamond prism allows for evanescent wave coupling to the resonator. In {type-I} parametric down-conversion, correlated photon pairs can be created in resonator modes phase matched with the pump. The intra-cavity pump power is chosen to be far below the OPO threshold determined in a previous experiment \cite{Furst:2010p6510}. The signal and the idler photons (as well as the residual pump) are coupled out of the triply-resonant WGMR using the same diamond prism and are separated by an additional dispersion prism. The idler photons are detected with one avalanche photodiode (APD)(blue). The signal photons are detected in a Hanbury Brown-Twiss setup, consisting of a 50:50 beamsplitter and two APDs (yellow). The residual pump is monitored with a PIN diode (red). (B) To verify coarse tuning of the signal and idler wavelengths, the output of the WGMR is analyzed with an optical spectrum  analyzer. (C) To verify fine tuning of the wavelength, we use a scanning Fabry-Perot resonator in the signal arm.}
\end{figure}

The heart of our experiment (Fig.~\ref{fig.setup}) is a crystalline WGMR with a high quality factor on the order of $Q \approx 10^{7}$ \cite{Furst:2010p1048}. The coupling of light to the resonator is realized via optical tunneling between a diamond prism and the resonator. With a triply-resonant configuration, we exploit natural phase-matching to achieve SPDC. We verify the spontaneous regime of the parametric process by demonstrating a linear dependence of the pair production rate on the pump power (see Supplementary Information). Correcting for the detection efficiency, we find a pair production rate of $1.3\cdot10^{7}$\,pairs/s per mW pump power with the bandwidth of 13 MHz. This efficiency is two orders of magnitude higher than the one found in previous experiments \cite{Scholz:2009p4886}.

The quantum state of the emitted photons can be characterized by observing the temporal correlations
of photon detection events between the signal and the idler arm, i.e. the Glauber
cross-correlation function $g_{si}^{(2)}$ \cite{Glauber:1963p6573} (see
Supplementary Information). For a Lorentzian line shape, one can calculate that $g_{si}^{(2)}(\tau)=1+\gamma/(2R)\cdot\exp(-\gamma|\tau|)$ \cite{Ou:1999p3986,Bettelli:2010p5282}, where $\gamma$
is the characteristic decay rate and $R$ is the pair production rate.

\begin{figure}
\begin{tabular}{l}
A\\[-0.3cm]
\centerline{\includegraphics[width=8.0cm]{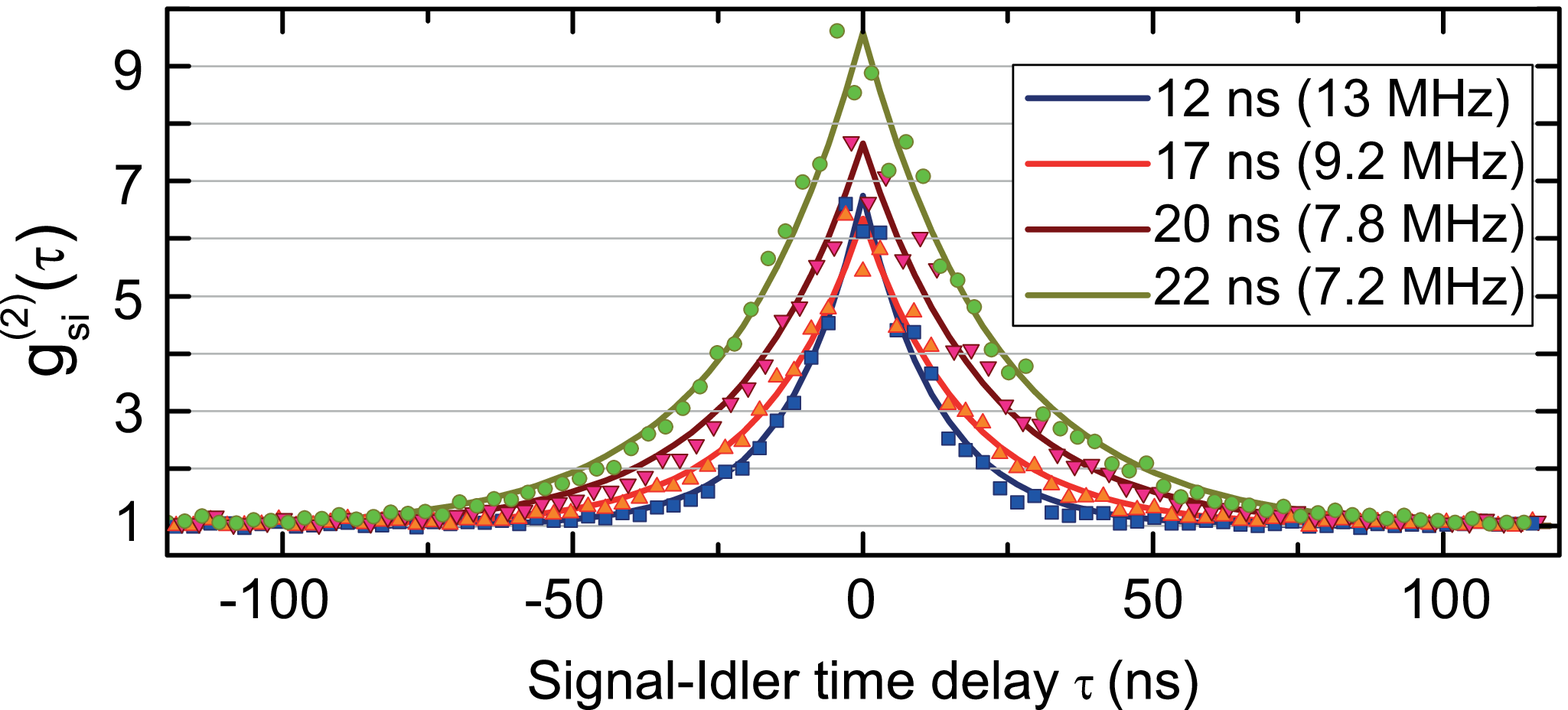}} \\
B\\[-0.3cm]
\centerline{\includegraphics[width=8.5cm]{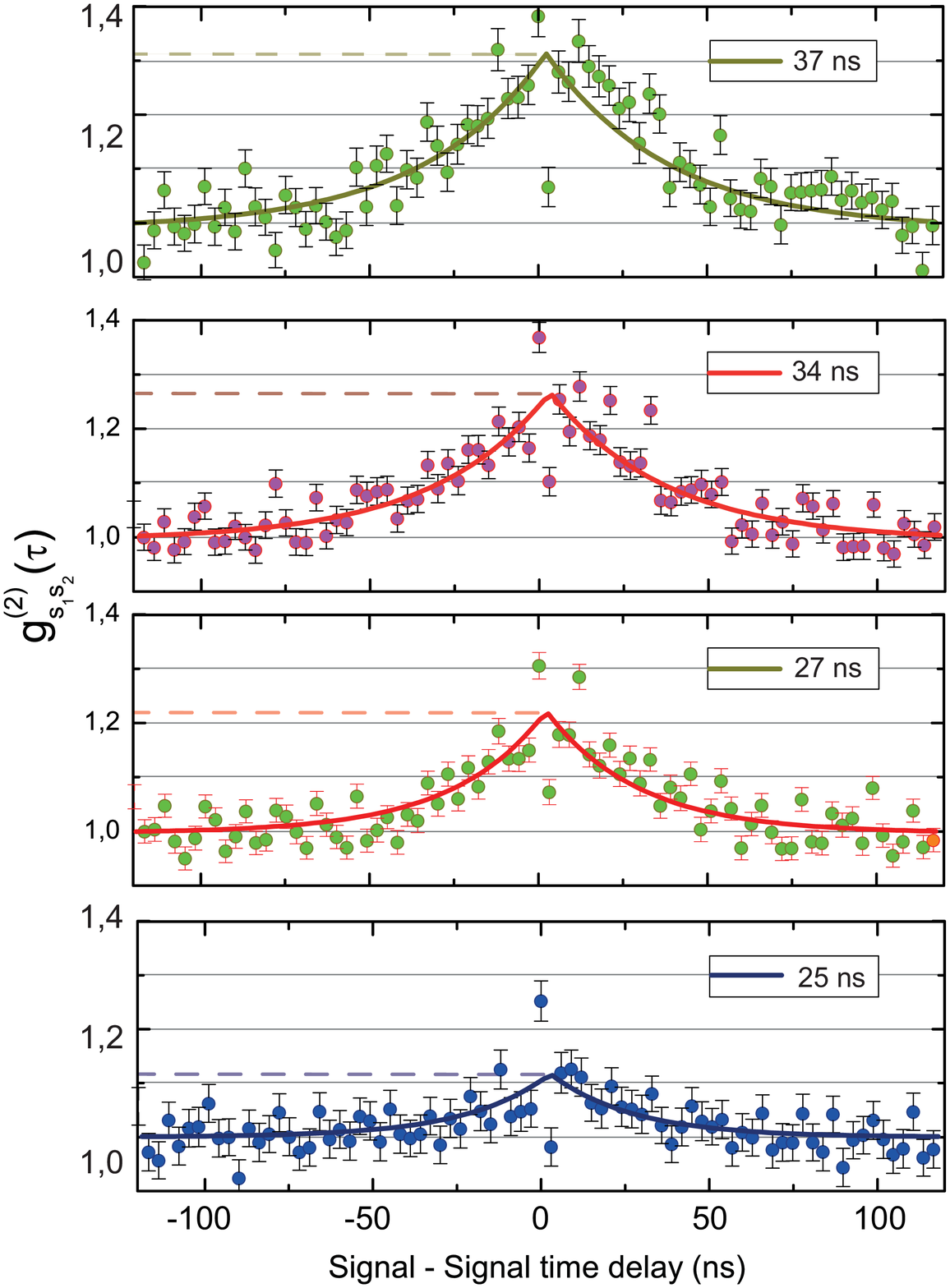}} \\
\end{tabular}
\caption{\label{fig.coincidences_all} (A) Normalized cross-correlation functions $g_{si}^{(2)}(\tau)$ measured for different coupling distances of the WGMR. The bandwidth of the generated photons is determined by fitting an exponential (solid lines) to each measured data set (dots). The exponential fits demonstrate a bandwidth tunability of nearly a factor of $2$, starting from the smallest experimentally verified bandwidth of $7.2\mbox{MHz}$. (B) Normalized autocorrelation function $g_{ss}^{(2)}(\tau)$ measured for the same coupling distances as in the case of the cross-correlation function (keeping color coding of Fig.~\ref{fig.coincidences_all}A).}
\end{figure}

For an SPDC process enhanced by a high-Q resonator, such as our WGMR, the width of the correlation function is determined by the coherence
time of the emitted photons corresponding to the bandwidth $\Delta\nu=\gamma/2\pi$ of the resonator. An exceptional feature of WGMRs is the widely and continuously tunable bandwidth $\Delta\nu$. This feature is accessible through the evanescent coupling, i.e. by varying the distance between the coupling prism and the resonator. Note that for conventional resonators, this would require a tunable transparency of the resonator mirrors which is very difficult to realize, especially for high-Q resonators. In Fig.~\ref{fig.coincidences_all}A, we present the experimental data for $g_{si}^{(2)}(\tau)$. Using exponential fits (solid lines) to the data (dots), we find the correlation time and determine the bandwidth $\Delta\nu$, acquired for different coupling distances. We demonstrate a tunability of bandwidth between $7.2\,\mbox{MHz}$ and $13\,\mbox{MHz}$, which can in principle be extended drastically (see Supplementary Information).

In a next step, the number of effective modes in the generated states is estimated from the normalized Glauber autocorrelation function $g_{s_{1}s_{2}}^{(2)}(\tau)$ \cite{Loudon:2001}. The value $g_{s_{1}s_{2}}^{(2)}(\tau=0)$ is expected to scale with the number of effective modes \cite{McNeil:1983p7129} $n$ as $1+1/n$. In Fig.~\ref{fig.coincidences_all}B, autocorrelation functions are shown for the same coupling distances as the cross-correlation functions in Fig.~\ref{fig.coincidences_all}A. The number of involved modes clearly depends on the distance between the prism and the resonator. 
By increasing this distance the peak values increases up to a maximum of $1.35$, which indicates three effective modes. We want to point out that this high quality is naturally achieved by our source without spectral and spatial filtering. In principle, efficient single-mode operation is feasible by optimizing the pump spectrum, and using even larger coupling distances or different resonator geometries, featuring a more sparse WGMR spectrum \cite{Ilchenko:2001p7395}.

For the experimental data in Fig.~\ref{fig.coincidences_all}B, the corresponding decay times are determined
by fitting exponentials to each data-set. 
We find that the decay times of the autocorrelation function exceed the decay times  of the cross-correlation function by a factor between 1.7 and 2.0. 
A difference in decay times is already expected for a single mode case \cite{Loudon:2001} and was observed in previous resonator assisted SPDC experiments \cite{Hockel:2011p4528}.

\begin{figure}
\begin{tabular}{l}
\centerline{\includegraphics[width=8.5cm]{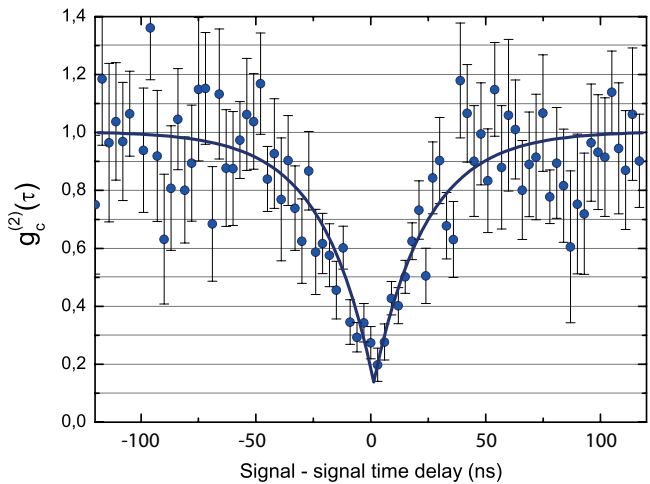}} \\
\end{tabular}
\caption{\label{fig.g2herald}Normalized Glauber autocorrelation function $g_{c}^{(2)}(\tau)$ conditioned on an idler detection event with a heralding interval of 20\,ns. The clearly visible anti-bunching dip proves the non-classicality of our heralded single
photon source.}
\end{figure}

In order to demonstrate heralded single photons, we evaluate the normalized Glauber autocorrelation function conditioned on an idler detection event $g_{c}^{(2)}(t_{s_{1}},t_{s_{2}}|t_{i})$ (for details see SOI). The minimum of the $g_{c}^{(2)}(t_{s_{1}},t_{s_{2}}|t_{i})$ function is sensitive to the nonclassical signal-idler correlations and tends to zero for heralded single photon states. In Fig.~\ref{fig.g2herald}, the special case of $g_{c}^{(2)}(\tau)$ 
is presented with a heralding interval of 20\,ns. The characteristic anti-bunching of $g^{(2)}(0) < 0.2$ is clearly visible, which reflects the non-classical correlations between two single photons in two different arms.

\begin{figure}
\begin{tabular}{l}
\\[0.3cm]
\centerline{\includegraphics[width=9cm]{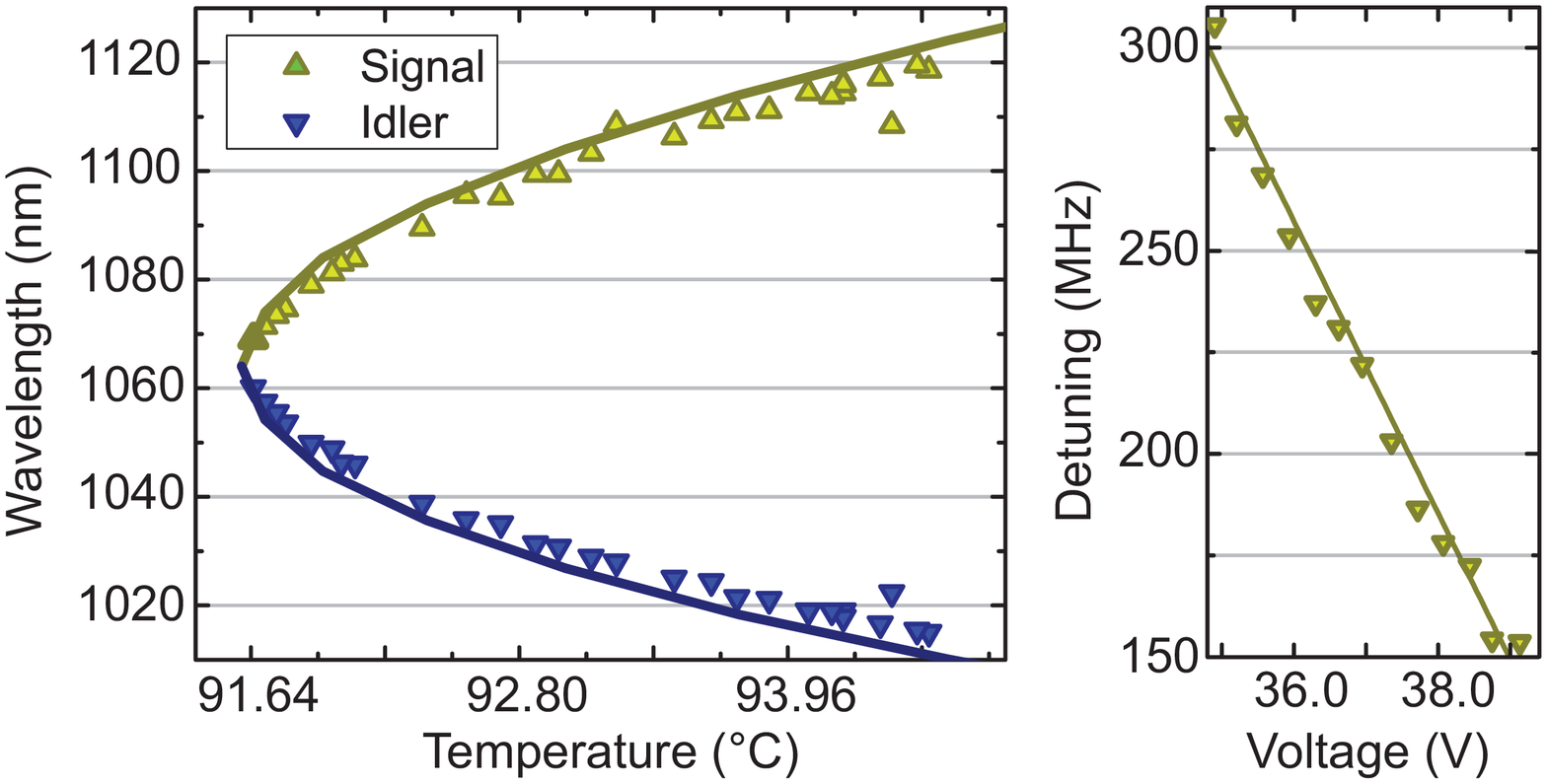}} \\[-4.8cm]
A\hspace{5.6cm}B\\ [4.5cm]
\end{tabular}\caption{\label{fig.bias_tuning} Wavelength tuning. (A) By changing the resonator temperature over $3^{\circ}\,\mbox{C}$, we measure a wavelength detuning of $100\,\mbox{nm}$ between signal and idler (triangles) with an optical spectrum analyzer. This agrees well with the phase matching condition for WGMRs (solid line). (B) By changing the bias voltage over a range of 4 V, we measure a mode-hop free linear wavelength detuning of $150\,\mbox{MHz}$ with a scanning Fabry-Perot interferometer (free spectral range of $1\,\mbox{GHz}$).
}
\end{figure}

For compatibility with various components such as stationary qubits (e.g. atomic transitions), non-classical light sources need to be tunable not only in bandwidth of the generated light but also in their central wavelength to match the qubit's transition. In a triply-resonant OPO, such wavelength tuning is realized by controlling the phase matching. In our setup, we have control over the optical length of the resonator via the applied temperature and the voltage. Because of the limited sensitivity of the wavelength measurement devices, the wavelength tuning demonstration was carried out above the OPO threshold.
Figure~\ref{fig.bias_tuning}A shows the wavelength tuning measured
with an optical spectrum analyzer while the resonator temperature
was changed (see Fig.~\ref{fig.setup}B). Varying the temperature
over $3^{\circ}\mbox{C}$ results in a wavelength detuning of $100\,\mbox{nm}$
between the signal and the idler. In Fig.~\ref{fig.bias_tuning}(B), we demonstrate a continuous fine-tuning of the central wavelength by applying a voltage across the WGMR.
This measurement is realized by sending the signal beam to a scanning
Fabry-Perot interferometer, as shown in Fig.~\ref{fig.setup}C. By
changing the bias voltage over a range of $4\,\mbox{V}$, we measure
a mode-hop free wavelength detuning of $150\,\mbox{MHz}$ which potentially
can be applied fast \cite{Guarino:2007p5605}.

Our results mark the starting point of a new class of resonator-enhanced SPDC sources, simultaneously easily tunable in bandwidth and wavelength while offering a compact, stable and easy to implement design with remarkable efficiency. Operating our setup in a Sagnac configuration can transform it to a polarization entanglement source \cite{Hentschel:2009p7405}. By changing the WGMR geometry and crystal doping and modifying the phase-matching temperature one is readily able to address the wavelength of various atomic transitions and connect them to the telecom wavelengths (see Supplementary Information). This renders the source ideal for the implementation of quantum repeaters and quantum information processing in general. 

We gratefully acknowledge the contribution of Gerhard Schunk, Malte Avenhaus and Andreas Christ and support from BMBF grant QuORep and EU grant QESSENCE.

\section{Supplementary Information}

\subsection{Experimental Details}
Our whispering gallery mode resonator (WGMR) is manufactured from a 5~\% MgO-doped lithium niobate $(\mbox{LiNbO}_{3})$ wafer. The symmetry axis of the resonator coincides with the crystal's optical axis. The system is similar to a triply resonant optical parametric oscillator (OPO) but is operated far below its experimentally determined pump power threshold \cite{Furst:2010p6510}. As a pump source, we use a continuous wave (cw) frequency doubled Nd:YAG laser (InnoLight, Prometheus) with a wavelength of $532~\mbox{nm}$ and a linewidth of 5~kHz. Typical pump powers at the input face of the coupling prism are on the order of several hundred nanowatts. The pump light is coupled to the resonator using a diamond prism and is coupled out together with the spontaneously generated signal and idler photons using the same prism. 
We match the pump laser to the WGMR by a sample and hold locking. During the 1~ms sample phase, we sweep the laser across the WGMR resonance and determine the resonance frequency by monitoring the pump transmission. During the 199~ms hold phase, the laser frequency is kept on the resonance.
Natural type I SPDC phase matching is realized by stabilizing the resonator temperature at the appropriate set point. An additional fast control over the effective resonator length is realized by a voltage applied across the resonator. Planar electrodes are placed on the top and bottom side of the disk resonator. 
Although we did not explicitly lock the parametric modes in the WGMR, the parametric down-conversion process was sufficiently stable to enable measurements for more than two weeks. Finally the signal and the idler are separated from each other and from the transmitted pump light using a dispersion prism and subsequently are directed to the detection stage. 

\subsection{Measurement of the correlation functions}
\label{meas_1}
In order to measure all three Glauber correlation functions, we use three avalanche photodiodes (APDs; id Quantique id400). The idler photons are detected with one APD whereas the signal photons are detected in a Hanbury Brown and Twiss (HBT) setup \cite{Brown:1956p7506} (two APDs placed after a $50:50$ beamsplitter). All APDs used in the experiment are operated in the Geiger mode and at a quantum efficiency of 7.5~\%. The electronic signals from the APDs are recorded by a time-to-digital converter (TDC; quTau, qutools GmbH) with a temporal resolution of $\tau_{d}=162 \mbox{ps}$. The TDC records incoming detection events from the APDs with their time-stamps and channel number and streams the acquired data to a computer. From the raw time-stamps the relative time delays were calculated using a software implemented start-stop algorithm. The bin size for all auto- and cross-correlation measurements was set to 3~ns. In order to avoid the influence of afterpulsing of the APDs on the statistics, we choose a slightly different dead time for each APD. The combination of low quantum efficiency and large deadtimes ($\approx 10 ~\mu\mbox{s}$) required to run the experiment for more than two weeks in order to present statistically significant results. Throughout the measurement period the system is operated without active stabilization of the distance between the coupling prism and the WGMR. To have a measure of the coupling distance, we continuously recorded the pump coupling efficiency, measured as the pump transmission contrast during the sample phase. In the undercoupled regime, a decreasing contrast corresponds to an increased distance.
 By the virtue of the fact that the idler and the transmitted pump light exit the WGMR through the same coupling prism, we can sort the data according to the WGMR coupling distances which determine the signal and idler bandwidths.

\subsection{Measurement of the wavelength tuning}
\label{meas_2}
Besides the possibility of tuning the bandwidth of the generated single photons, our system has the ability to tune their central wavelengths. 
The effective resonator lengths for the involved modes are dependent on temperature and voltage, e. g. due to thermo-refractive and electro-optical effects. This gives us control over the phase matched wavelength triplet in the PDC process. The wavelength tuning based on the resonator bias voltage allows for a rapid and continuous (mode-hop free) tuning within over a hundred MHz. Changing the resonator temperature allows for a slow tuning of each signal and idler over several tens of nanometers. In order to verify the wavelength tuning in the regimes mentioned, we use a scanning Fabry Perot resonator (FPR) and an optical spectrum analyzer (OSA). Because of the low sensitivity of these instrument's sensors, we measured the signal and idler beams when operating above the OPO threshold. The tunability demonstration nonetheless remains valid because the phase-matching conditions do not depend on sub- or above threshold operation.

\subsubsection{Tuning based on variation of the resonator bias voltage}
We place a scanning FPR with a free spectral range (FSR) of $1~\mbox{GHz}$ in the signal beam. While continuously modulating the bias voltage, we monitor the transmission through the FPR on an oscilloscope and find a tunability of the center wavelength in the MHz regime. The FPR is scanned over more than one FSR to calibrate the measurement results.

\subsubsection{Tuning based on variation of the temperature}
For the measurement of the wavelength tuning based on variation of the temperature, the dispersion prism is substituted by a polarization beamsplitter. Consequently, only the vertically polarized pump is separated from the horizontally polarized parametric beams. Signal and idler are jointly coupled to an optical multimode fiber and guided to the OSA, which allows us to confirm the tunability over several tens of nanometers. For all temperatures the pump laser is coupled to the same resonator mode, although with changing frequency.

\subsection{Linear pump conversion}
\label{linear_conversion}
\begin{figure}
\begin{tabular}{l}
\\[0.3cm]
\centerline{\includegraphics[width=8.4cm]{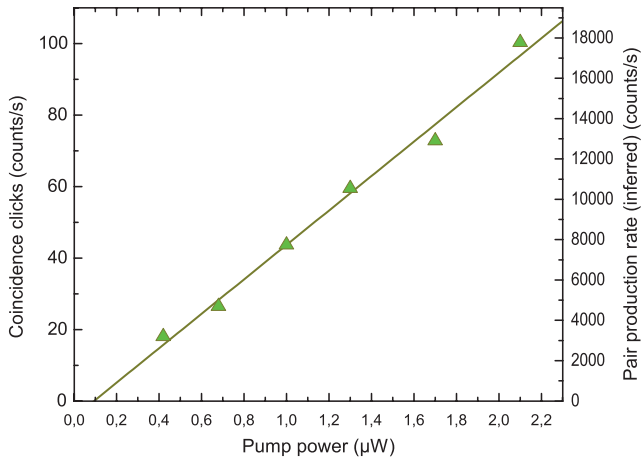}} \\\end{tabular}
\caption{\label{fig.pump_conversion}The coincidence clicks and the pair production rate depending on the pump power (each measurement time 30~min, coincidence window 30~ns).}
\end{figure}

Far below the OPO threshold, one expects the rate of emitted photon pairs to scale linearly with the pump power. 
This indicates that the parametric gain is low. Therefore, on average we have much less than one photon per coherence volume, i.e. per mode. To verify this in our experiment, we measured the coincidence rate for different pump powers. The coincidence window for these measurements was set to 30~ns. The data was collected for 30~min for different pump powers incident on the coupling prism. The distance between the coupling prism and the resonator was set to the same value for all measurements.
The results presented in Fig. \ref{fig.pump_conversion} demonstrate a clear linear dependence between the pair production rate and the pump power. The pair production rate is inferred from the coincidence rate by taking into account the APD's quantum efficiencies ($\eta=7.5\,\%$ for each APD). 

\subsection{Normalized cross-correlation function}
\label{function explanation}

\begin{figure}
\begin{tabular}{l}
\\[0.3cm]
\centerline{\includegraphics[width=8.4cm,height=7.4cm]{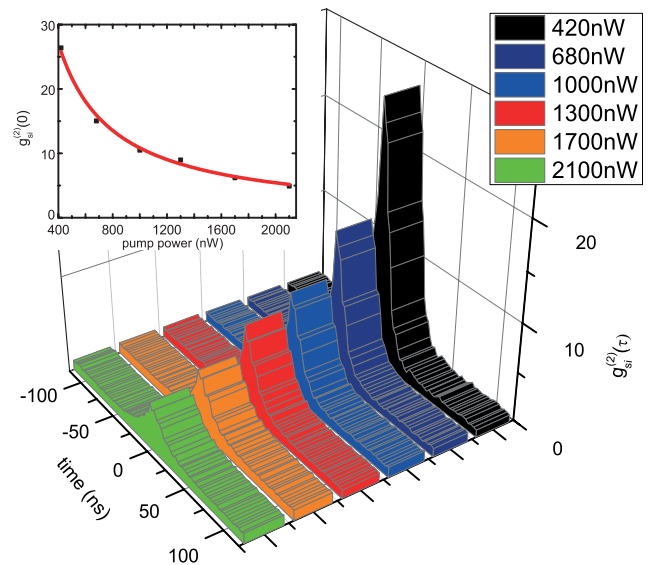}} \\\end{tabular}
\caption{\label{fig.pump_rate_inset}The $g_{si}^{(2)}(\tau)$ function for different pump powers (equal coupling conditions, each measurement time 30 min, binning time 3~ns). The inset shows $g_{si}^{(2)}(\tau=0)$ as a function of the pump power which is proportional to the pair production rate.}
\end{figure}

One way to characterize the quantum state of the generated stream of photons is to analyze the temporal correlations between the signal and idler light~\cite{Kimble:1977p3991,Walls:1979p7559}, i.e. the normalized Glauber inter-beam (cross-)correlation function \cite{Glauber:1963p6573,Loudon:2001}:
\begin{equation}
\label{eq:coinc}
g_{si}^{(2)}(\tau) = 
\frac{\langle E_{s}^{\dagger}(t + \tau) E_{i}^{\dagger}(t) E_{i}^{}(t) E_{s}^{}(t + \tau)\rangle}
{\langle E_{s}^{\dagger}(t + \tau) E_{s}^{}(t + \tau)\rangle \langle E_{s}^{\dagger}(t) E_{s}^{}(t)\rangle}, 
\end{equation}
where $E_{s}^{}(t)$ and $E_{i}^{}(t)$ represent the scalar positive-frequency field operators for the outgoing signal (s) and idler (i) beams and the averaging is done over time \cite{Wong:2006p7507}.
In general, the cross-correlation function reflects the temporal correlations between the signal and the idler photons and is sharply peaked around zero with a width of $\Delta t /2$, where $\Delta t$ is the signal-idler correlation time. The signal-idler peak reflects the pair-generation rate and could be made arbitrary large by reducing the intensity of the pump \cite{Bettelli:2010p5282}. We verified that the peak height of the $g_{si}^{(2)}(\tau)$ is inversely proportional to the pump power (Fig. \ref{fig.pump_rate_inset}).  

In a resonator-enhanced spontaneous parametric down-conversion (SPDC) process, the coherence time is determined by the resonator's ring-down time which is tantamount to the bandwidth $\Delta \nu$ of the resonator's mode.  Since high-Q resonators, such as our WGMR, have ring-down times of several tens of nanoseconds, the detector response, in the order of hundreds of picoseconds, and the binning time of 3~ns play a negligible role.

\subsection{Normalized Glauber autocorrelation function}
\label{signal_signal_autocorrelation}
The normalized autocorrelation functions is defined as
\begin{equation}
\label{eq:signal-signal}
g_{s_1 s_2}^{(2)}(\tau) = 
\frac{\langle E_{s_1}^{\dagger}(t + \tau) E_{s_2}^{\dagger}(t) E_{s_2}^{}(t) E_{s_1}^{}(t + \tau)\rangle}
{\langle E_{s_1}^{\dagger}(t + \tau) E_{s_1}^{}(t + \tau)\rangle \langle E_{s_2}^{\dagger}(t) E_{s_2}^{}(t)\rangle},
\end{equation}
where the subscripts $s_1$ and $s_2$ indicate the single photons detected at two ports of the 50:50 beamsplitter. The maximum of this function is expected to scale with $1+1/n$, where $n$ represents the number of effective modes. We introduce the term "effective modes", since this scaling law assumes that the modes are equally excited. This might not be the case in our experiment.

The difference in the decay times between the cross- and the autocorrelation function can be understood by considering the spectral properties of the two-mode squeezing operator. This operator is determined by the Hamiltonian that governs the system's dynamics. For a Gaussian spectrum this difference has already been investigated in \cite{Loudon:2001}. A detailed analysis of the spectral properties of the SPDC process including the spectral properties of the pump and the relation between the two widths in such systems will be provided elsewhere. 

\subsection{Conditioned normalized Glauber autocorrelation function}
\label{idler_triggered_autocorrelation}
Due to the fact that an OPO intrinsically is not a single photon source, it is important to characterize the generated states in terms of photon number distributions. Here the capability of always generating two photons at the same time is used to gain information about the generated states by measuring the normalized Glauber autocorrelation function of the signal channel conditioned on an idler detection event \cite{Bettelli:2010p5282}
\begin{equation}
\label{eq:heraldg2_2}
g_{c}^{(2)}(t_{s_1},t_{s_2}|t_{i}) =
\frac{P_{s_1 s_2,t_i}(t_{s_1}^{} ,t_{s_2}^{},t_i)}
{R^{3}_{} g_{si}^{(2)}(t_{s_1}^{}-t_i) g_{si}^{(2)}(t_{s_2}^{} - t_i)},
\end{equation}
where $P_{s_1 s_2,t_i}$ is the triple-coincidence rate and $R$ is the pair-generation rate. Signal photons in either detector $s_1$ and $s_2$ are considered for the heralding condition.
Because of the limited detection efficiency of $\eta ^2 = 5.6 \cdot 10^{-3}$ and long dead times of approximately $10\, \mu s$ of the APDs, we introduce a heralding interval $[-\tau_{\mbox{h}}, \tau_{\mbox{h}}]$ for accepting the arrival time of a signal photon triggered on an idler event. The width of the heralding interval is chosen to compensate for the small detection probability but simultaneously be significantly smaller than the coherence time of the detected photons. The presented data is the result of a stable measurement carried out over more than two weeks. We choose a heralding time of $\tau_{\mbox{h}} = 10 ~\mbox{ns}$, which is almost a factor 5 smaller than the corresponding coherence time of the detected photons. In the experimental measurement of $g_{c}^{(2)}$ the minimum value can be influenced by stray light and residual infrared light from the frequency doubled pump laser.

\subsection{Bandwidth tuning}
\label{bandwidthtuning}
A perfect resonator (without any losses) would confine light for an infinite amount of time corresponding to resonant modes at precise frequencies. The important parameter describing any deviations from this ideal case is the resonator's Q(uaility) - factor. In general, each loss can be described by an individual Q-factor defined by the central frequency $\nu$ and the corresponding loss rate $\Delta \nu_{i}$: 
\begin{equation}
\label{quality_factor}
\mbox{Q}_{i} = \frac{\nu}{\Delta \nu_{i}}
\end{equation}
The total Q-factor of a resonator is defined as the inverse sum of all individual Q-factors and is directly connected with the bandwidth of the resonator $\Delta \nu$:
\begin{equation}
\label{quality_factor_1}
\mbox{Q}^{-1} = \sum_{i}\mbox{Q}_{i}^{-1}
\end{equation}
\begin{equation}
\label{quality_factor_2}
\mbox{Q} = \frac{\nu}{\Delta \nu}
\end{equation}
Since we are using a polished, macroscopic WGMR, the dominant losses in our system are coupling and intrinsic absorption. 

The Q-factor for the intrinsic absorption is determined by the material of the resonator, i.e. by the materials absorption coefficient $\alpha$ and its refractive index $n_{s}$:
\begin{equation}
\label{absorption_q_factor}
\mbox{Q}_{a} = \frac{2 \pi n_{s}}{\alpha \lambda}.
\end{equation}

The equation for the coupling Q-factor, when using prism coupling, can be derived from \cite{Gorodetsky:1999p1550} as:
\begin{equation}
\label{coupling_q_factor}
\mbox{Q}_{c} = \frac{\sqrt{2} \pi^{5/2} n_{s}^{1/2} (n_{s}-1)}{(n_{c}^{2} - n_{s}^{2})^{1/2}} 
(\frac{a}{\lambda})^{3/2} \exp{(2 \gamma d)},
\end{equation}
where $n_{c}$ is the refractive index of the prism, $a$ is the radius of the resonator, $\lambda$ is the wavelength, $\gamma = \sqrt{k^2 (n_{s}^{2} - 2)}$, and $d$ is the distance between the prism and the resonator.

\begin{figure}
\begin{tabular}{l}
\\[0.3cm]
\centerline{\includegraphics[width=8.4cm]{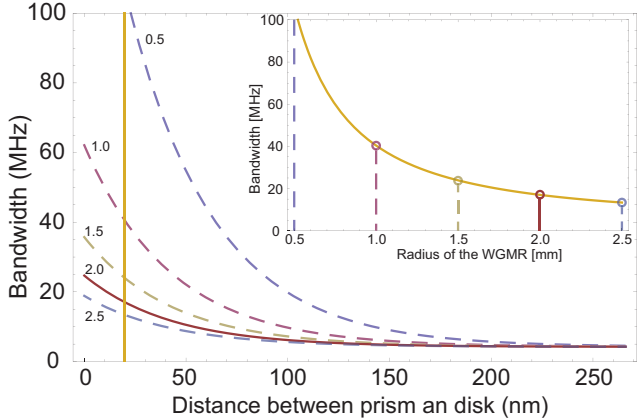}} \\\end{tabular}
\caption{\label{fig.bandwidth_tuning}The bandwidth of a resonator mode versus the distance between the coupling prism and the resonator illustrated for different WGMR radii and a center wavelength of 1064~nm. Inset: Dependency of the bandwidth on the WGMR radius for a fixed gap distance of 20~nm.}
\end{figure}

Here one of the advantages of WGMRs becomes clearly visible since the ratio between the intrinsic Q-factor and the coupling Q-factor is not fixed but can continuously be modified by changing the distance $d$ between the prism and the resonator. The resulting change in the total Q-factor finally leads to the desired bandwidth tunability of the resonant mode. Changing the resonator bandwidth affects the conversion efficiency. In Fig. \ref{fig.bandwidth_tuning} the solid purple line represents the bandwidth tunability of our WGMR  for a center wavelength of $1064~\mbox{nm}$. Moreover, by changing the radius of the resonator one can influence the dynamic range of the bandwidth tunability which is represented by the 4 dashed lines in Fig. \ref{fig.bandwidth_tuning}. The inset of Fig. \ref{fig.bandwidth_tuning} highlights the dependency of the dynamic range on the resonator radius at fixed gap distance of $20~\mbox{nm}$. 

\subsection{Wavelength tuning}
\label{wavelength_tuning}
\begin{figure}
\begin{tabular}{l}
\\[0.3cm]
\centerline{\includegraphics[width=8.4cm]{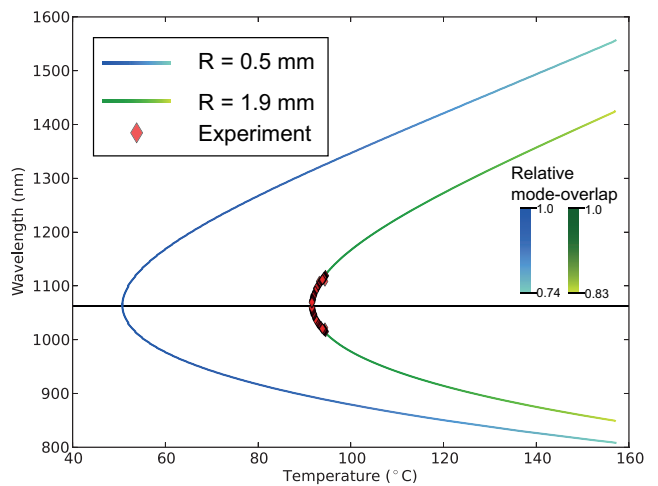}} \\\end{tabular}
\caption{\label{fig.wavelength_tuning_2}Wavelengths of signal and idler versus the phase-matching temperature for resonators with a radius of 1.9~mm (green) and 0.5~mm (blue).}
\end{figure}

For the theoretical prediction of the coarse wavelength tuning we calculate the wavelengths for signal $\lambda_s$ and idler $\lambda_i$ which fulfill the type I phase matching condition for our pump wavelength $\lambda_p$ \cite{Kozyreff:2008p1056}. This condition strongly depends on the bulk refractive index \cite{Schlarb:1994p7580}, on the shape of the resonator, on the temperature, and on the transversal mode structure \cite{Gorodetsky:2006p6331}. Fig. \ref{fig.wavelength_tuning_2} shows the numerical solutions for two resonators with different radii. 

In general the phase-matching function consists of continuous segments with a slope of 5~MHz/mK and discontinuous steps of one free spectral range, i.e. mode hops. The relative insensitivity of the continuous segments on temperature changes is one major reason for the stability of your system. The overall temperature depended wavelength tuning is realized by mode hops over several free spectral ranges of the resonator. To bridge the frequency gaps between individual mode hops one uses non-equatorial conversion channels and the electro-optical effect. 

The green line in Fig. \ref{fig.wavelength_tuning_2} represents the solution for a resonator with a  radius of 1.9~mm and is in good agreement with our experimental results (red rhombi). 
Due to technical reasons we did not exceed the temperature level of $94^{\circ}$~C.
The blue line shows the numerical solution for a resonator with a radius of 0.5~mm and demonstrate that by decreasing the resonator radius the whole phase-matching is shifted towards lower temperatures. For both theoretical predictions the relative mode-overlap is individually color-coded and normalized to the maximum of each phase-matching curve. For this theoretical predictions we only considered the phase matching between fundamental resonator modes.
\newpage

\bibliography{wgm_arXiv_3}
\end{document}